\documentclass[12pt]{iopart}

\usepackage{graphicx}
\usepackage{cite}
\usepackage{color}
\usepackage{multirow}

\begin{document}

\setlength{\parindent}{0pt}

\title[Bistable solutions for the electron energy distribution function]{Bistable solutions for the electron energy distribution function in electron swarms in xenon: a comparison between the results of first-principles particle simulations and conventional Boltzmann equation analysis}

\author{Nikolay Dyatko$^1$ and Zolt\'an Donk\'o$^2$}

\address{$^1$State Research Center of Russian Federation Troitsk Institute for Innovation and Fusion Research, Troitsk, Moscow
142190, Russia\\
$^2$Institute for Solid State Physics and Optics, Wigner Research Centre for Physics, Hungarian Academy of Sciences, 1121 Budapest, Konkoly Thege Mikl\'os str. 29-33, Hungary.\\ }
\ead{dyatko@triniti.ru, donko.zoltan@wigner.mta.hu}

\begin{abstract}
At low reduced electric fields the electron energy distribution function in heavy noble gases can take two distinct shapes. This ``bistability effect'' -- in which electron-electron (Coulomb) collisions play an essential role -- is analyzed here for Xe with a Boltzmann equation approach and with a first principles particle simulation method. The solution of the Boltzmann equation adopts the usual approximations of (i) searching for the distribution function in the form of two terms (``two-term approximation''), (ii) neglecting the Coulomb part of the collision integral for the anisotropic part of the distribution function, (iii) treating Coulomb collisions as binary events, and (iv) truncating the range of the electron-electron interaction beyond a characteristic distance. The particle-based simulation method avoids these approximations: the many-body interactions within the electron gas with a true (un-truncated) Coulomb potential are described by a Molecular Dynamics algorithm, while the collisions between electrons and the background gas atoms are treated with Monte Carlo simulation. We find a good general agreement between the results of the two techniques, which confirms, to a certain extent, the approximations used in the solution of the Boltzmann equation. The differences observed between the results are believed to originate from these approximations and from the presence of statistical noise in the particle simulations.
\end{abstract}


\submitto{\PSST}
\maketitle

\section{Introduction}

The physics of electron swarms has been attracting considerable attention during the past decades because of the interest in the various physical effects taking place in these settings and by the need for accurate input data in discharge modeling \cite{lxcat,lxcat2}. The advance of the theoretical background, as well as of the computational resources and numerical techniques made it possible to develop a detailed picture of the physics of particle swarms. Most of the efforts have been devoted to electron swarms, the electron energy distribution function and the transport properties have been determined for a wide variety of gases and gas mixtures, and for a broad range of conditions, see e.g.  \cite{Florian,Loff,Robson,Zoran09,Dujko11,Pinhao,Trunec,White,DY,Sasa,Nap}. 

One of the particular phenomena, the {\it bistability of the electron energy distribution function} (EEDF) is a pronounced manifestation of the nonlinear nature of these systems. The term ``EEDF  bistability'' is used here to designate a situation when in a physical system, under fixed conditions (gas number density, gas temperature, electric field strength, electron and ion concentrations, population of excited states), two stable steady states are possible with different EEDFs. 

The electron temperature bistability was (according to our best knowledge) predicted for the first time for a plasma under the effect of an alternating electromagnetic field, at conditions when electron-ion collisions are dominant \cite{int1,int2}. Assuming a Maxwellian EEDF, the bistability in the plasma of heavy rare gases (argon, krypton and xenon) with an applied electric field, was shown to exist in \cite{int3} due to the specific shapes of the momentum transfer cross sections for the electron scattering from Ar, Kr and Xe atoms. Later, the effect in heavy rare gases plasma was confirmed by using a more accurate approach, the Boltzmann equation (BE) analysis that took into account electron-electron (e-e) collisions \cite{int4,int5,int6}. In \cite{int4,int5,int6} it was found that, within a certain range of parameters: (i) the reduced electric field strength, $E/n$ and (ii) the electron to neutral atom density, $n_{\rm e}/n$ (where $E$ is the electric field strength, $n$ is the atom number density, and $n_{\rm e}$ is the electron density), the BE has two stable solutions. In the case of xenon gas, e.g., the  bistability effect was found at $n_{\rm e}/n \geq 10^{-9}$ and 0.025 Td $< E/n <$ 0.043 Td. 

The conditions mentioned above can be realized in different physical systems, e.g., in decaying plasmas in the presence of low electric field, in non-self-sustained discharges, and under electron swarm conditions. A decaying plasma in Xe was studied in Ref. \cite{int3}, where the time-dependence of the current in the afterglow plasma in the presence of a weak  electric field was measured. A jump-like decrease in the current was observed at some time point during the plasma decay, which was attributed to the manifestation of the bistability effect \cite{int3}. Non-self-sustained discharges (sustained by a beam of fast electrons) were considered in Refs. \cite{int7,int8}, in which the possibility of the existence of the EEDF bistability was theoretically analyzed in Ar, Kr and Xe. It was shown that the  bistability effect takes place in Xe and Kr, while for an Ar plasma \cite{int8}, the BE has a unique solution over the entire parameter range examined. To our best knowledge, the bistability effect has not been investigated so far for electron swarm conditions. 

We note that besides pure noble gases, the bistability effect has also been studied in gas mixtures. In \cite{int9} the electron temperature ($T_{\rm e}$) in Ar:N$_2$ = 100:1 mixture afterglow was studied both experimentally and theoretically. In the experiments it was observed that, under certain conditions, a rather sharp knee appears in $T_{\rm e}(t)$, while the vibrational temperature of nitrogen molecules, $T_{\rm v}$, remains almost constant. The theoretical study of the EEDF in Ar:N$_2$ afterglow plasma was carried out by the numerical solution of the appropriate BE, by taking into account e-e collisions, as well as superelastic vibrational and superelastic electronic collisions. The calculations have shown that ranges of $n_{\rm e}$ and $T_{\rm v}$ exist, where two different solutions of BE can be obtained. The observed knee-like time-dependence of $T_{\rm e}$ was explained as the manifestation of the bistability effect. The possibility of this effect in pure nitrogen afterglow plasma was theoretically addressed in \cite{int10}. In addition, using the Maxwellian distribution function approach the  bistability effect was studied in \cite{int11} for positron swarms in He. For more details the reader is referred to the review \cite{review}.

It should be noted that so far the theoretical analysis of the bistability effect was performed based on BE approach only.  In these studies, and generally, in the solutions of the Boltzmann equation including electron-electron collisions, rather serious approximations have traditionally been used: the methods (i) search for the distribution function in the form of two terms (``two-term approximation''), (ii) neglect the electron-electron part of the collision integral for the anisotropic part of the distribution function, (iii) treat Coulomb collisions as binary events, and (iv) truncate the range of the electron-electron interaction beyond a characteristic distance. These approximations became widely accepted throughout the years, and one needs to note that in the absence of a more rigorous approach, their effects on the results of BE solutions have never been critically examined. To do that - and this is actually the motivation of our work - one would ideally describe the physical system of interest with a {\it first-principles method that does not involve any approximations}. (We note that previous particle-based (Monte Carlo) methods, dealing with Coulomb collisions, are neither free of approximations.) The particle-based simulation technique, used here \cite{donko}, on the other hand, relaxes both the above major approximations and provides a solution based on first principles. Therefore we expect this method to be able to evaluate the accuracy of the BE solution that is limited by the approximations adopted, and can verify the effects predicted within the frame of the BE approach, like the EEDF bistability that was introduced above and is studied here in details in Xe gas.

The physical system investigated here can be described by three independent parameters: the reduced electric field, $E/n$, the electron to gas (number) density ratio, $n_e/n$, and the electron density, $n_e$. In the description of the solution of the Boltzmann equation (section 3.2) this will be shown to follow from the structures of the equations involved. Equivalently, other sets of parameters derived from the above set can be used as well: we shall carry out our studies by varying $E/n$ and $n_e/n$, while keeping the gas number density, $n$, constant. 

In section 2 we introduce the underlying physical processes responsible for the development of the bistability effect. This is followed in section 3 by the presentation of the methods of calculation: section 3.1 discusses the particle simulation method, while section 3.2 outlines the solution of the Boltzmann equation. The results obtained by the two different approaches are presented and compared to each other in section 4. A short summary is given in section 5.

\section{Qualitative explanation of the EEDF bistability effect in rare gases}

The qualitative explanation of the possibility of the EEDF  bistability effect in rare gases at weak electric fields (where only elastic collisions occur) adopts an approximation of a Maxwellian EEDF (i.e., it assumes {\it a priori} that the EEDF is Maxwellian, $f_0(\varepsilon) \sim \exp(-\varepsilon/ k_{\rm B} T_{\rm e})$, for details see e.g. \cite{review}). For such conditions, the time-dependent equation for the electron temperature is as follows:
\begin{equation}
\frac{{\rm d}T_{\rm e}}{{\rm d}t} = n ~\Phi(E/n,T_{\rm e}) = n~ \bigl[ H(E/n,T_{\rm e}) - L(T_{\rm e},T_{\rm g}) \bigr],
\label{eq:balance}
\end{equation}
where the term $H(T_{\rm e})$ describes the heating of electrons by the electric field and the term $L(T_{\rm e})$ accounts for the loss of electron energy in elastic collisions. (Note that, generally, the loss rate is positive, as electrons deposit energy in collisions with the background gas. However, at $T_{\rm e} < T_{\rm g}$, where $T_{\rm g}$ is the gas temperature, electrons are heated by the atoms of the gas, resulting in a negative loss, $L < 0$.) 

\begin{figure}[ht]
\begin{center}
\includegraphics[width =0.9\textwidth]{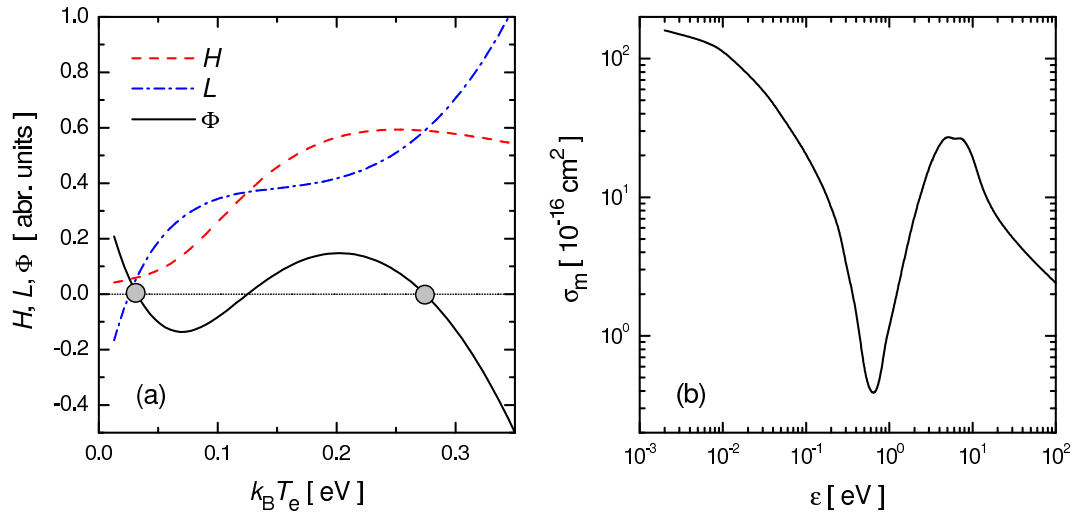}
\caption{(a) Heating ($H$) and loss ($L$) rates of the electron energy, due to the effect of the electric field and to their interaction with the background gas atoms, respectively. The curves have been derived assuming a Maxwellian EEDF and a gas temperature of $T_{\rm g}$ = 300 K. The data are given in arbitrary units, the grey dots indicate the stable equilibrium solutions, with $\Phi=H-L=0$. (Note that a similar plot appeared in Ref. \cite{review}, as figure 27. In that figure the labelling of the $H$ and $L$ curves was exchanged as a misprint.) (b) Elastic momentum transfer cross section for electron -- Xe atom collisions \cite{Park}.}
\label{fig:1}
\end{center}
\end{figure}

The number of steady state and stable solutions of (\ref{eq:balance}) depends on the shape of the $\Phi(T_{\rm e})$ function. Figure \ref{fig:1}(a) shows this function together with the  $H(T_{\rm e})$ and $L(T_{\rm e})$ functions \cite{review}, for Xe at $E/n$ = 0.035 Td, and a gas temperature of $T_{\rm g}$ = 300 K. Let us note that $L(T_{\rm e})$ is proportional to the elastic momentum transfer cross section $\sigma_{\rm m}(\varepsilon)$ of electron scattering from Xe atoms, while $H(T_{\rm e})$ varies proportionally to $1/\sigma_{\rm m}(\varepsilon)$ (see section 3a). Due to the specific shape of $\sigma_{\rm m}(\varepsilon)$ (a deep Ramsauer-Townsend minimum at an energy of $\approx$ 0.6 eV, see figure \ref{fig:1}(b)) the $H(T_{\rm e})$ and $L(T_{\rm e})$ curves exhibit a ``wavy'' behavior and cross each other at three points. As a consequence, the $\Phi(T_{\rm e})$ function has an ``inverse-N-type'' shape and is equal to zero at three $T_{\rm e}$ values. Among these three steady state solutions of equation (\ref{eq:balance}) two (where $\Phi(T_{\rm e})$ has a negative slope) are stable, as indicated in figure \ref{fig:1}(a).

Although the EEDF in reality is generally not Maxwellian, but the e-e collisions drive the EEDF towards Maxwellian, therefore the above explanation holds at least qualitatively. We note that e-e collisions are essential in establishing the bistability effect, as the terms corresponding to these collisions introduce non-linearity into the BE. Without e-e collisions the BE is linear equation (in $f_0$). As a linear equation, it has a unique solution at fixed parameters (taking into the account the proper normalization and the condition that $f_0 \rightarrow 0$ at electron energy $\rightarrow \infty$). If e-e collisions are involved, the BE becomes a non-linear equation and, in principle, two or more different solutions are possible. For multiple solutions to appear the e-e collision frequency should be high enough to have an influence on the shape of the EEDF, i.e., it should be comparable or higher than the frequency of energy losses in electron-atom elastic collisions. We note, however, that the non-linearity is a necessary, but not a sufficient condition for BE to have multiple solutions. While some examples of physical systems, where bistability is present, were given in section 1, a general sufficient condition for the appearance of the effect cannot be formulated at present.  

\section{Methods}

Here we describe the basics of the two computational approaches, the particle simulation scheme is presented in section 3.1, while the solution of the Boltzmann equation is discussed in section 3.2.

\subsection{Particle simulation}

A swarm of electrons is simulated under the effect of a homogeneous electric field, in xenon gas, at different values of the number density ratio, $\eta = n_e / n$. At the low reduced fields considered only elastic collisions take place. As this ensures conservation of the number of particles, we consider a spatially homogeneous system. 

The simulation scheme is based on a combination of a Molecular Dynamics (MD) technique and a Monte Carlo (MC) approach \cite{donko}. The MD describes the many-body interactions (driven by the inter-particle Coulomb potential) within the classical electron gas, while the MC part handles the interaction of the electron gas with the background (atomic) gas. 

\subsubsection{Molecular Dynamics simulation of the electron gas.}

In our classical many-body system the particles move under the influence of interparticle and external forces in a cubic simulation box, with periodic boundary conditions. The equations of motion of particles $i = 1,\dots,N$, which are numerically integrated with discrete time steps ($\Delta t$), are Newtonian:
\begin{equation}
m \frac{{\rm d}^2 {\bf r}_i}{{\rm d}t^2} = \sum_{i \neq j} {\bf F}_{ij} - e {\bf E},
\end{equation}
where $m$ and $-e$ are the mass and the charge of the electron, respectively. The sum on the right hand side represents the force exerted on particle $i$ by all other ($j \neq i$) particles {\it and their periodic images} located in spatial replicas of the simulation box. These images have to be included in the proper determination of the interparticle forces in the case of the ``full'' (un-truncated) infinite-range Coulomb potential. (Note that for our conditions no screening of the potential takes place by oppositely charged species.) This summation is indeed a key issue and needs a special approach, as will be explained below. The second term on the right hand side of the above equation is a contribution due to the external electric field. We note that in the absence of the interaction of the electrons with a background gas the electrons would continuously be accelerated by this field, which will, however, not be the case when e$^-$+Xe atom collisions take place. 

The MD method adopted in this study is based on the Particle-Particle Particle-Mesh (PPPM) approach, described in details in \cite{HE}, that uses a partitioning of the interaction into (i) a force component that can be calculated on a mesh (the ``mesh force'') and (ii) a short-range (``correction'') force, which is to be applied to closely separated pairs of particles only. In the mesh part of the calculation, charge clouds are used instead of point-like charges. The charge density distribution $\rho({\bf r})$ is assigned to a grid and is Fourier transformed to the ${\bf k}$-space. Multiplying $\rho({\bf k})$ with an optimized Green function results in a potential distribution $\phi({\bf k}) = G({\bf k})\rho({\bf k})$, which is subsequently transformed back to real space. The forces acting on the particles are obtained by differentiation of the potential and interpolating the electric field to the positions of the particles. The cloud shape is chosen in a way to ensure that $\rho({\bf k})$ is band-limited (which would not hold for point-like charges). The calculation of the potential in the Fourier space automatically takes into account the periodic images of the primary computational box. For closely separated particles a correction force is to be applied, which is the difference between the forces between two point-like charges and two charges with the cloud shape used. The simulations describe a micro-canonical ensemble, where the number of particles, the volume of the system and energy are conserved.

The upper limit for the simulation time step is defined by the stability of the integration of the equation of motion in the case of the closest approach of two electrons, $r_{\rm min} = e^2/(4 \pi \epsilon_0 \varepsilon_{\rm max})$. Here $\varepsilon_{\rm max}$ is a pre-defined maximum energy \cite{HE}, which has to be chosen carefully, to ensure that the probability of finding electrons with $\varepsilon > \varepsilon_{\rm max}$ is vanishingly small at the conditions considered. In this work we adopt a maximum energy of 5 eV, for this value the time step has to be chosen to be as low as $\Delta t \cong 2.2 \times 10^{-16}$ s. This results in a very demanding computational load to follow the evolution of the system for a long time. As a converged solution assumes sufficient interaction between the electrons and the gas as well, one needs to have a high-enough electron-atom collision frequency. This can only be ensured by setting a high the gas atom number density. For this reason our computations are carried out with $n = 2.5 \times 10^{27}$ m$^{-3}$, which allows simulated times of tens of nanoseconds (with a computational speed of simulating $\sim$ 1 ns in one day, on a single CPU). The number of electrons is chosen to be $N$ = 1000.

\subsubsection{Monte Carlo simulation of electron gas - background gas interaction.}

Having solved the description of the electron gas with the MD method that accounts for electron-electron interactions, now we introduce a background gas and let the two gases to interact via e$^-$+Xe collisions. The probability of an e$^-$+Xe collision during a time step $\Delta t$ is calculated as:
\begin{equation}
P_{\rm coll} = 1 - {\rm exp}[-n \sigma_{\rm m}(g) g \Delta t],
\end{equation}
where $g = |{\bf g}|$, with ${\bf g}= {\bf v} - {\bf V}$ being the relative velocity between the electron and a Xe atom with a velocity ${\bf V}$ randomly chosen from a Maxwellian background of gas atoms having a temperature $T_{\rm g}$. This probability is calculated for each electron in each time step, and decision about the occurrence of a collision is made by comparing $P_{\rm coll}$ with a random number. Collisions are executed in the center-of-mass frame, and are considered to be isotropic. The energy change of the gas atoms colliding with the electrons is not accounted for, the gas temperature is kept constant.

The simulations allow investigation of both the stationary state and the transients (induced by changing the electric field strength or the gas temperature). The transients are followed by monitoring the time-dependence of the mean energy of the electrons, which is calculated in each time step. The EEDF is, on the other hand, only calculated for the stationary state as its accurate determination (over 6--7 orders of magnitude) requires averaging over millions of time steps. 

\subsection{Solution of the Boltzmann equation}

Here we discuss the specific features of the BE for electrons, as applied to conditions under consideration (homogeneous plasma, or an electron swarm in an atomic gas, acted upon by a weak steady electric field). The distribution function of the electrons, $f({\bf v})$, can be described by the equation 
\begin{equation}
\frac{\partial f({\bf v})}{\partial t} - \frac{e {\bf E}}{m} ~ \nabla_{\bf v} f({\bf v}) =C,
\label{eq:be1}
\end{equation}
where {\bf v} is the electron velocity and $C$ is the collision integral. Further we shall consider only elastic scattering of electrons from atoms and electron-electron collisions: $C = C_{\rm m} + C_{\rm e}$.

The solution of the BE is based on expansion of the distribution function in Legendre polynomials $P_n(\cos \Theta)$, in which only two first terms are taken into account:
\begin{equation}
f({\bf v})= f_0(v) + f_1(v) \cos \Theta, 
\label{eq:be2} 						
\end{equation}
where $v$ is the velocity magnitude, $\Theta$ is the angle between ${\bf v}$ and $-{\bf E}$, $f_0(v)$ is the symmetrical part of the distribution function and $f_1(v)$ describes the directed motion of the electrons along the electric field. (We note that this ``two-term approximation'', in the absence of e-e collisions, has been benchmarked with other solution methods of the BE, as well as with particle-based (Monte Carlo) simulations in several studies, see e.g. \cite{Nuno}.)

The substitution of expansion (\ref{eq:be2}) into equation (\ref{eq:be1}) leads to equations for the $f_0$ and $f_1$ functions:
\begin{equation}
\frac{\partial f_0}{\partial t} - \frac{e E}{3 m v^2} \frac{\partial}{\partial v} (v^2 f_1) = C_{\rm 0m} + C_{\rm 0e} 
\label{eq:f0}
\end{equation}
and
\begin{equation}
\frac{\partial f_1}{\partial t}- \frac{e E}{m} \frac{\partial f_0}{\partial v} = C_{\rm 1m} + C_{\rm 1e}. 
\label{eq:f1}
\end{equation}

The collision integrals $C_{\rm 0m}$ and $C_{\rm 1m}$ can be written as \cite{int2}:
\begin{eqnarray}
C_{\rm 0m} =  \frac{1}{2 v^2} \frac{\partial}{\partial v} \Biggl[ \frac{2m}{M} \nu_{\rm m} v^2 \biggl( \frac{k_{\rm B}T_{\rm g}}{m} \frac{\partial f_0}{\partial v} + v f_0 \biggr) \Biggr], \\
C_{\rm 1m} = -\nu_{\rm m} f_1, 
\end{eqnarray}
where $\nu_{\rm m} = n \sigma_{\rm m} v$ is the momentum transfer frequency and $\delta = 2m/M$ is the average fraction of the energy lost by the electrons in one elastic collision with atom ($M$ is the mass of the gas atom). The rate of the electron energy loss due to elastic collisions is characterized by the frequency $\nu_{\rm u} = \delta \nu_{\rm m}$.

It is known that the calculation of the pair-collision frequency in the case of Coulomb collisions encounters a characteristic difficulty, namely the logarithmic divergence of frequency at small scattering angles. This difficulty is avoided by assuming that the Coulomb potential acts only up to a certain finite distance $r_{\rm max}$ (see later). The expression for the term $S_{\rm 0e}$ is written as follows \cite{int2,SJB}:
\begin{eqnarray}
S_{\rm 0e} = \frac{1}{v^2} \frac{\partial}{\partial v} \Biggl\{ v^2 \nu_{\rm e} \biggl[ A_1(f_0) v f_0 + A_2 (f_0) \frac{\partial f_0}{\partial v} \biggr]  \Biggr\}, \\
A_1(f_0) = 4 \pi \int_0^v v_1^2 f_0(v_1) {\rm d}v_1,\\
A_2(f_0) = \frac{4 \pi}{3} \Biggl[ \int_0^v v_1^4 f_0(v_1) {\rm d}v_1 + v^3 \int_v^\infty v_1 f_0(v_1) {\rm d}v_1 \Biggr],
\end{eqnarray}
where $f_0(v)$ is normalized as:
\begin{equation}
4 \pi \int_0^\infty v^2 f_0(v) {\rm d}v = 1
\end{equation}
and 
\begin{equation}
\nu_{\rm e} = 2 \pi n_e \Biggl( \frac{e^2}{4 \pi \varepsilon_0 m}\Biggr) \frac{1}{v^3} \ln \Biggl[ 1 + \biggl(\frac{r_{\rm max}}{r_0}\biggr)^2 \Biggr]~~,~~
r_0 = \frac{e^2}{4 \pi \varepsilon_0 m v^2}. 
\label{eq:nue}
\end{equation}

In BE calculations the value of the parameter $r_0$ is usually estimated by means of the mean electron energy, that gives $r_0 = e^2 / 4 \pi \varepsilon_0 2 \overline{\varepsilon}$ . For the case of plasmas the Debye length, $\lambda_{\rm D}$, is taken to be the cutoff distance, i.e. $r_{\rm max}=\lambda_{\rm D}$. For the case of swarm conditions considered here we use the approximation:\begin{equation}
r_{\rm max} = 0.5 n_{\rm e}^{-1/3},
\label{eq:rmax}
\end{equation}
i.e. the half of the average distance between the electrons is taken to be the cutoff distance. (This choice of $r_{\rm max} $ is based on intuitive physical considerations: if the impact parameter (of test electron relative to a given electron) is higher than the half the average distance between electrons in the gas, then the influence of this electron on the test one becomes weaker than the influence of the neighboring one.) Note that in most cases $r_{\rm max}/r_0 \gg 1$ (see comments in \cite{int2,SJB}) and the ``1'' in the expression under the logarithm in (\ref{eq:nue}) can be omitted. 

As to the term $S_{\rm 1e}$ in equation (4), it is very complex (see comments in \cite{int2}) and we did not find publications in which this term was taken into account in calculations. As a rule (BOLSIG+  \cite{r4}, EEDF \cite{r5,r6}), it is neglected assuming that $\nu_{\rm e} \ll \nu_{\rm m}$. Then, if the characteristic time of plasma parameters variation is essentially smaller than $\nu_{\rm m}^{-1}$, the time derivative in equation (\ref{eq:f1}) can be omitted. In this case  
\begin{equation}
f_1  = \frac{e E}{\nu_m m} \frac{\partial f_0}{\partial v}
\label{eq:f1final}
\end{equation}
and equation for the $f_0$ function is written as
\begin{equation}
\frac{\partial f_0}{\partial t} - \frac{e E}{3 m v^2} \frac{\partial}{\partial v} \Biggl( v^2 \frac{e E}{\nu_{\rm m} m} \frac{\partial f_0}{\partial v} \Biggr) = C_{\rm 0m} + C_{\rm 0e}.
\label{eq:be17}
\end{equation}

It should be noted that, in the absence of e-e collisions, the parameter for the steady state solution of equation (\ref{eq:be17}) is the reduced electric field $E/n$. If the e-e collisions are taken into account, as it has already been mentioned in section 1, there are three parameters: $E/n$, $n_e/n$ and $n_e$. The electron number density is an independent parameter since the logarithmic term in eq. (\ref{eq:nue}) (the Coulomb logarithm) depends on $n_e$. Actually, at fixed $E/n$ and $n_e/n$ values the dependence of $f_0$ on $n_e$ is rather weak, since $n_e$ value is under logarithm.

For the numerical solution of (\ref{eq:f1final}) it is rewritten with energy as a variable. The steady-state equation is solved by an iteration method similar to that in \cite{r6,r7}. The initial $f_0(\varepsilon)$ is assumed to be the Maxwellian with a given electron temperature $T_{\rm e0}$. In the case of calculation of the time-dependent solution of the BE the time step should be as small as $\Delta t \ll \nu_{\rm u}^{-1}$ and $\Delta t \ll \nu_{\rm e}^{-1}$. For conditions under consideration $\Delta t$ was taken as $\Delta t = 2 \times 10^{-14}$ s.  

\section{Results}

\begin{figure}[ht]
\begin{center}
\includegraphics[width =0.9\textwidth]{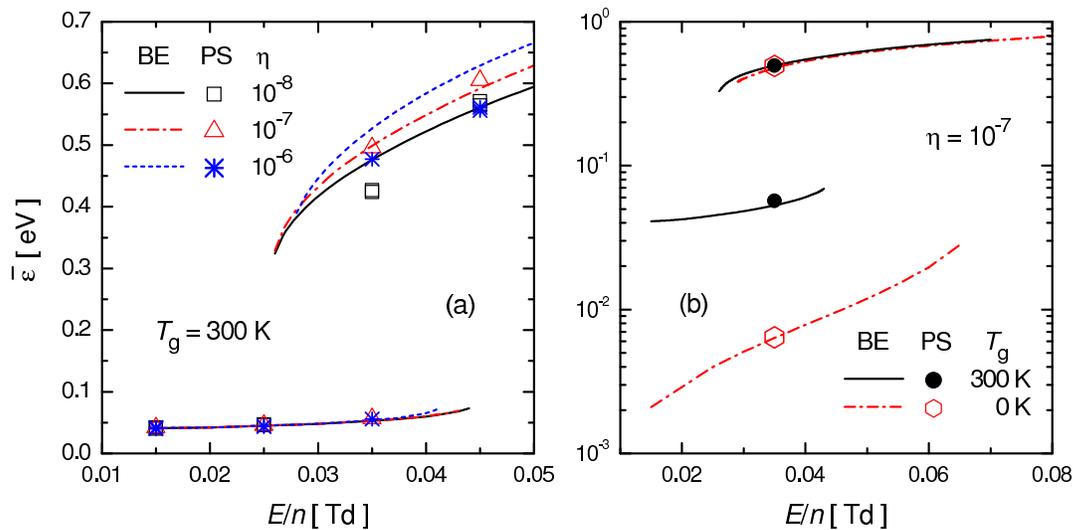}
\caption{(a) Mean energy values obtained from the calculations at different values of the electron to gas density ratio, $\eta$. The results of both the Boltzmann equation solutions (``BE'', lines) and of particle simulations (``PS'', symbols) originate from calculations started with different initial conditions. $T_{\rm g}$ = 300 K. (b) The effect of the gas temperature ($T_{\rm g}$) on the limits of the bistable solutions, at $\eta = 10^{-7}$. }
\label{fig:2}
\end{center}
\end{figure}

Bistability of the EEDF, i.e. two solutions for the distribution function, for certain conditions are found here, both via the solution of the Boltzmann equation and via executing the particle-based simulations. To find these different solutions both methods start from different initial conditions: both assume a Maxwellian distribution with a ``low'' and a ``high'' mean energy. For the particle method $\bar{\varepsilon}_{\rm init} = $ 0.009 eV and 0.9 eV are used as ``low-energy'' and ``high-energy'' initial conditions, with random initial positions of the particles in the simulation box. For the BE solution the initial mean energy values are: $\bar{\varepsilon}_{\rm init}$ = 0.065 eV and 0.65 eV. In the domain where a unique solution exists the results of the calculations do not depend on these choices. In the case of two possible solutions two domains of $\bar{\varepsilon}_{\rm init}$ exist: from one of these domains ($\bar{\varepsilon}_{\rm init} < \bar{\varepsilon}_{\rm init}^\ast$) the calculations converge to the lower energy solution, and from the other domain ($\bar{\varepsilon}_{\rm init} > \bar{\varepsilon}_{\rm init}^\ast$) to the higher energy solution. The value of $\bar{\varepsilon}_{\rm init}$ within these domains has no effect on the results. The boundary value, $\bar{\varepsilon}_{\rm init}^\ast$, depends on the $E/n$ and $\eta$.
  
\begin{figure}[h!]
\begin{center}
\includegraphics[width =1\textwidth]{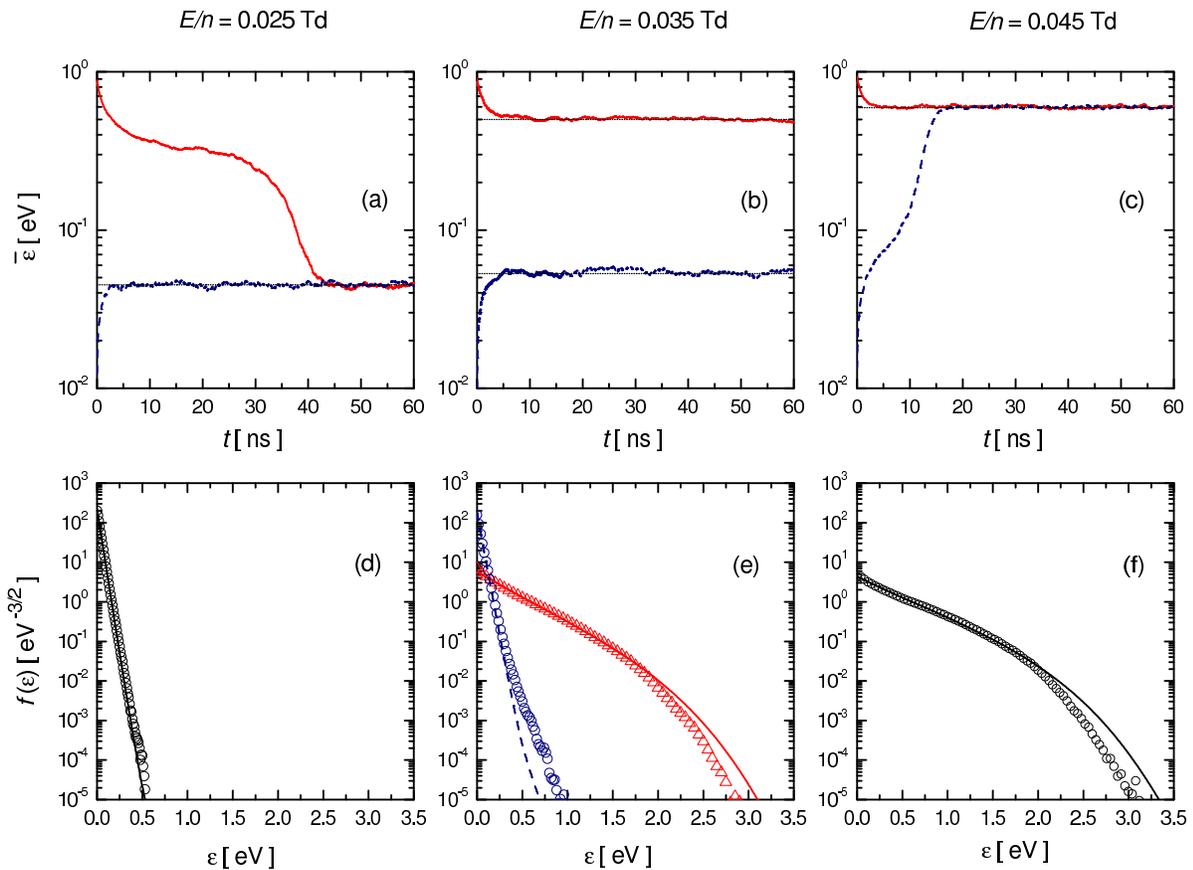}
\caption{(a,b,c) The convergence of the particle simulation method starting from a ``high-energy'' distribution (solid red lines) and a ``low-energy'' distribution (dashed dark blue lines), for the $E/n$ values indicated. Two solutions are found for $E/n$ = 0.035 Td. The dotted horizontal black lines indicate the mean energy values obtained from BE calculations. (d,e,f) $f(\varepsilon)$ for the stable solutions: Boltzmann equation data (lines) in comparison with particle simulation data (symbols). $\eta=10^{-7}$ for all plots.}
\label{fig:3}
\end{center}
\end{figure}  
  
The BE analysis has been carried out for a range of reduced electric fields, 0.015 Td $\leq E/n \leq$ 0.05 Td, for various electron to gas density ratios. Results of these calculations, in terms of the mean electron energy $\bar{\varepsilon}$, for $\eta = 10^{-8}, 10^{-7}$, and $10^{-6}$ are displayed in figure \ref{fig:2}(a). Two solutions are found for the intermediate values of $E/n$, the boundaries of the domain change slightly with the density ratio. The ``low-energy'' solution exhibits a mean energy that is almost independent of $\eta$, and is in the range $\bar{\varepsilon} \sim 0.04 \dots 0.05$ eV. In the case of the ``high-energy'' solutions the mean energy amounts several tenth of an electron Volt and increases with increasing $\eta$. The particle simulation results agree generally well with those obtained from the BE, small differences are found in the case of ``high-energy'' solutions. Additionally, the particle simulation does not predict a bistability at $\eta=10^{-8}$ at $E/n$= 0.035 Td, whereas the BE solution does. The reason of this discrepancy is not fully understood. We contemplate that the presence of noise in the particle simulation, due to its statistical nature, prevents finding solutions where the minimum in the energy balance vs. $T_{\rm e}$ is very shallow. This, however, needs further clarification. While all the data in figure \ref{fig:2}(a) have been shown for a gas temperature of $T_{\rm g}$ = 300 K, figure \ref{fig:2}(c) shows the dependence of the results on $T_{\rm g}$, for fixed $\eta=10^{-7}$. The mean energy here is displayed on a log scale, as the gas temperature influences predominantly the ``low-energy'' solutions. For a lower $T_{\rm g}$ a lower $\bar{\varepsilon}$ is found, however, the domain of $E/n$, where bistability is found, is wider at $T_{\rm g}$ = 0 K. 

The stationary solutions in the particle simulations were typically obtained beyond a few tens of nanoseconds of simulated time. The convergence of the mean energy is illustrated in figures \ref{fig:3}(a)-(c), for reduced electric fields of $E/n$ = 0.025 Td, 0.035 Td, and 0.045 Td, respectively, at $\eta=10^{-7}$. In all cases the simulations were started from two initial configurations, as already mentioned above. While at $E/n$ = 0.025 Td and 0.045 Td we observe convergence to a unique value of $\bar{\varepsilon}$, the runs at $E/n$ = 0.035 Td clearly yield two solutions, with mean electron energies differing by almost an order of magnitude. The full EEDFs, obtained by the two methods, are compared in figures \ref{fig:3}(d)-(f). The EEDFs are normalized as $\int_0^\infty f(\varepsilon) \sqrt{\varepsilon} {\rm d}\varepsilon =1$. We find a good agreement between the data obtained from the BE and from the particle simulation, although some differences show up in the tails of the distribution functions. The difference of the EEDFs belonging to the two stable solutions at $E/n$ = 0.035 Td is remarkable.

\begin{figure}[ht]
\begin{center}
\includegraphics[width =0.5\textwidth]{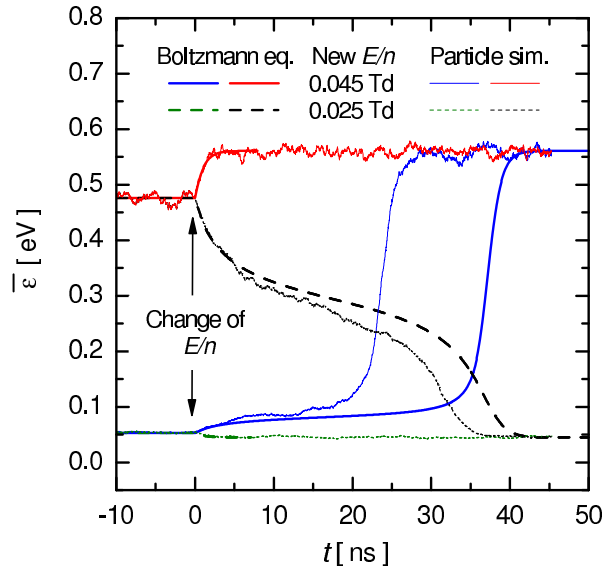}
\caption{Time dependence of the mean electron energy (response of the swarm) following a sudden change of the electric field (at $t=0$) from $E/n$ = 0.035 Td to $E'/n$ = 0.025 Td and 0.045 Td. The thick lines correspond to solutions of the BE, while thin lines show the particle simulation results. Unique solutions are found for the ``new'' values of the reduced electric field. $\eta=10^{-6}$}
\label{fig:4}
\end{center}
\end{figure}

To demonstrate the ability of our methods to follow the temporal evolution of the system here we take as an example the change of the electric field. We start from the two stationary stable solutions obtained at $E/n$ = 0.035 Td (at $\eta=10^{-6}$), change $E/n$ to a ``new'' value, and monitor the convergence of $\bar{\varepsilon}(t)$. We test ``new'' values of $E/n$  = 0.025 Td and 0.045 Td. The results obtained by both methods are depicted in \ref{fig:4}. Following the change of the reduced field both methods -- although with a different dynamics -- show the relaxation of the system to unique states, corresponding to the new values of $E/n$. The relaxation time is found to be of the order of few times ten nanoseconds. Here we recall the importance of the energy loss frequency $\nu_{\rm u}$, as this defines the time scale of the relaxation of the mean energy. An estimation at $\varepsilon$ = 0.5 eV gives $\nu_{\rm u}^{-1} \approx$ 20 ns, in accordance with our observations of the relaxation time scale.

The peculiarities of the transitions can be understood by following the heating and the cooling of the electrons at the specific conditions. The rate of electron heating is proportional to $E^2/\sigma_{\rm m}(\varepsilon)$ and the rate of energy loss is proportional to $\sigma_{\rm u}(\varepsilon)~ \varepsilon  =(2m/M)\sigma_{\rm m}(\varepsilon) ~\varepsilon$. 

First we discuss the transition from the ``low-energy'' solution at $E/n$ = 0.035 Td to the final state at $E/n$ = 0.045 Td. The change of the mean energy in this case is slow at the beginning but becomes very fast afterwards. In the initial state the EEDF is narrow and the mean energy is low, where the momentum transfer cross section is high. As the heating rate is proportional to $\sigma^{-1}_{\rm m}(\varepsilon)$ and the cooling rate is proportional to $\sigma_{\rm m}(\varepsilon)$, the rate of change of $\bar{\varepsilon}$ is low at high $\sigma_{\rm m}$ in the first phase of the transition. When the mean energy becomes higher $\sigma_{\rm m}(\varepsilon)$ decreases, giving more preference to the heating. This, together with the increasing slope of $\sigma_{\rm m}(\varepsilon)$, results in an abrupt increase of $\bar{\varepsilon}$ beyond a certain time. 

The transition from the ``high-energy'' solution at $E/n$ = 0.035 Td to the final state at $E/n$ = 0.025 Td is more complex. The three stages of relaxation -- seen in figure \ref{fig:4} -- can be explained as follows. Under the steady state conditions at $E/n$ = 0.035 Td the mean electron energy is relatively high, $\bar{\varepsilon} \approx $ 0.48 eV. At this energy the momentum transfer cross section is low (due to the Ramsauer-Townsend minimum) and, as a consequence, the rate of electron heating is relatively high. The high rate of heating is balanced by a high rate of energy loss. The decrease of the electric field causes an instant decrease of the heating. The unbalanced high rate of energy loss leads to the rapid decrease in the mean electron energy (during the first stage of $\varepsilon(t)$ relaxation) and, consequently, to a decrease of the rate of energy loss. The decrease in the rate of energy loss leads to near-balanced conditions, i.e. the difference between the rate of energy loss and the rate of heating becomes moderate, leading to the flattening of the $\varepsilon(t)$ curve (second stage of relaxation). When the mean energy decreases below $\varepsilon \approx$ 0.2 eV, the momentum transfer cross section increases sharply and the rate of energy loss increases noticeably, while the rate of heating decreases significantly. Consequently, $\varepsilon$ decreases faster below 0.2 eV, during the third stage of the relaxation. We note that since under the conditions considered here the electron-electron collision frequency is higher than the electron-atom energy exchange frequency ($\nu_{\rm u}$), the EEDF is nearly Maxwellian during the relaxation process. 
 
The transitions, which involve a small change of the mean energy (from the ``high-energy'' solution at $E/n$ = 0.035 Td to the final state at $E/n$ = 0.045 Td, and from the ``low-energy'' solution at $E/n$ = 0.035 Td to the final state at $E/n$ = 0.025 Td) are much faster, compered to the two cases discussed above.

\section{Summary}

We have investigated the bistability of the EEDF in Xe gas at low reduced electric fields via the solution of the Boltzmann equation and via a first principles particle simulation technique. The solution of the Boltzmann equation adopted the usual, widely accepted approximations: it (i) searched for the distribution function in the form of two terms, (ii) neglected the electron-electron part of the collision integral for the anisotropic part of the distribution function, (iii) treated Coulomb collisions as binary events, and (iv) truncated the range of the electron-electron interaction beyond a characteristic distance. The particle simulation method \cite{donko}, being devoid of any of these approximation has provided first-principles solutions to the problem, via a combination of a Molecular Dynamics simulation method (that described accurately the many-body interactions within the electron gas governed by the full Coulomb potential) and a Monte Carlo method (that handled the interaction of the electrons with the atoms of the background gas). Both methods allowed the computation of the EEDF and the related quantities, and have indicated the existence of two stable solutions for the EEDF for a range of $E/n$. The electron mean energies and the full EEDFs, obtained by the two methods agreed generally well for most od the parameter settings covered. Differences found for the domain of the bistability and for the shapes of the EEDFs may be attributed to approximations adopted in the BE solutions and to the presence of statistical noise in the particle simulations.

\ack The authors thank Prof. A. P. Napartovich for useful discussions and acknowledge the supported provided by the Hungarian Fund for Scientific Research (OTKA), via grant K 105476.


\section*{References}
\begin{thebibliography}{99}

\bibitem{lxcat} Pancheshnyi S, Biagi S, Bordage M C, Hagelaar G J M,  Morgan W L, Phelps A V, Pitchford L C 2012 {\it Chemical Physics} {\bf 398} 148

\bibitem{lxcat2} Alves L L, Bartschat K, Biagi S F, Bordage M C, Pitchford L C, Ferreira1 C M, Hagelaar G J M, Morgan W L, Pancheshnyi S, Phelps A V, Puech V and Zatsarinny O 2013 {\it J. Phys. D: Appl. Phys.} {\bf 46} 334002

\bibitem{Florian} Robson R E, Winkler R Sigeneger F 2002 {\it Phys. Rev. E} {\bf 65} 056410 

\bibitem{Loff} Loffhagen D, Braglia G L and Winkler R 2006 {\it Contrib. Plasma Phys.} {\bf 38} 527

\bibitem{Robson} Robson R E, Nicoletopoulos P, Li B and White R D 2008 {\it Plasma Sources Sci. Technol.} {\bf 17} 024020

\bibitem{Zoran09} Petrovi\'c Z Lj, Dujko S, Mari\'c D, Malovi\'c G, Nikitovi\'c Z, \v{S}a\v{s}i\'c O, Jovanovi\'c O, Stojanovi\'c V and Radmilovi\'c-Radenovi\'c M 2009 {\it J. Phys. D: Appl. Phys.} {\bf 42} 194002

\bibitem{Dujko11} Dujko S, White R D, Petrovi\'c Z Lj and Robson R E 2011 {\it Plasma Sources Sci. Technol.} {\bf 20} 024013 

\bibitem{Pinhao} Pinh\~ao N R, Donk\'o Z, Loffhagen D, Pinheiro M and Richley E A 2004 {\it Plasma Sources Sci. Technol.} {\bf 13} 719  

\bibitem{Trunec} Trunec D, Bonaventura Z and Ne\v{c}as D 2005 {\it J. Phys. D: Appl. Phys.} {\bf 39} 2544 

\bibitem{White} White R D, Brunger M J, Garland N A, Robson R E, Ness K F, Garcia G, de Urquijo J, Dujko S and Petrovi\'c Z Lj 2014 {\it European Physical Journal D} {bf 68} 125

\bibitem{DY} Deng Yunkun, Lu Chengdong and Xiao Dengming 2012 {IEEE Trans. Plasma Sci.} {\bf 40} 2671

\bibitem{Sasa} Dujko S, Raspopovi\'{c}, White R D, Makabe T and Petrovi\'c Z Lj 2014 {\it Eur. Phys. J. D} {\bf 68} 166

\bibitem{Nap} Napartovich A P and Kochetov I V 2011 {\it Plasma Sources Sci. Technol.} {\bf 20} 025001

\bibitem{int1} Gurevich A V 1958 {\it JETP} {\bf 35} 392; 1959 {\it Soviet Phys. JETP} {\bf 8} 271 

\bibitem{int2} Ginzburg V L and Gurevich A V 1960 {\it Soviet Physics Uspekhi} {\bf 3} 115; 1960 {\it Usp. Fiz. Nauk} {\bf 70} 201

\bibitem{int3} Gerasimov G N, Maleshin M N and Petrov S Ya 1985 {\it Opt. Specrosc. (USSR)} {\bf 59} 562

\bibitem{int4} Ivanov V A and Prikhod'ko A S 1986 {\it Sov. Phys.-Tech. Phys.} {\bf 31} 1202 

\bibitem{int5} Dyatko N A and Napartovich A P 2002 {\it Proceedings of 15th ESCAMPIG (Grenoble, France, 2002)} {\bf 1} 215

\bibitem{int6} Dyatko N A, Ionikh Y Z, Meshchanov A V and Napartovich A P 2006 {\it AIP Conference Proceedings} {\bf 876} 15

\bibitem{int7} 	Dyatko N A and Napartovich A P 2003 {\it J. Phys. D: Appl. Phys.} {\bf 36} 2096

\bibitem{int8} 	Dyatko N A and Napartovich A P 2004 {\it Plasma Physics Reports} {\bf 30} 953

\bibitem{int9} 	Dyatko N A, Ionikh Yu Z, Kolokolov N B, Meschanov A V and Napartovich A P 2000 {\it J. Phys. D: Appl. Phys.} {\bf 33} 2010

\bibitem{int10} Dyatko N A, Kochetov I V and Napartovich A P 2004 {\it Plasma Physics Reports} {\bf 30} 953

\bibitem{int11} Robson R E 1986 {\it J. Chem. Phys.} {\bf 85} 4486 

\bibitem{review} Dyatko N A, Kochetev I V, Napartovivh A P 2014 {\it Plasma Sources Sci. Technol.} {\bf 23} 043001

\bibitem{donko} Donk\'o Z 2014 {\it Phys. Plasmas} {\bf 21} 043504

\bibitem{Park} Park J L, Voshall R E, Phelps A V and Kline L E 1992 {\it J. Appl. Phys.} {\bf 71} 5363

\bibitem{HE} R. W. Hockney and J. W. Eastwood 1981 {\it Computer Simulation Using Particles} (New York: McGraw-Hill)

%\bibitem{r2} Landau L D and Lifshitz E M 1958 {\it Mechanics} (Gostekhizdat)

%\bibitem{r3} Rabinovich R I 1978 {\it Sov. Phys. JETP} {\bf 48} 262

\bibitem{Nuno} Pinh\~ao N R, Donk\'o Z, Loffhagen D, Pinheiro M J and Richley E A 2004 {\it Plasma Sources Sci. Technol.} {\bf 13} 719

\bibitem{SJB} Shkarofsky I P, Jonston T W and Bachynski M P 1966 {\it The particle kinetics of plasmas} (Addison-Wesley Publishing Company)

\bibitem{r4} Hagelaar G J M and Pitchford L C  2005 {\it Plasma Sources Sci. Technol.} {\bf 14}  722

\bibitem{r5} Dyatko N A, Kochetov I V, Napartovich A P and Sukharev A G {\it EEDF: the software package for calculations of the electron energy distribution function in gas mixtures} http://www.lxcat.laplace.univ-tlse.fr/software/EEDF/ 

\bibitem{r6} Dyatko N A, Kochetov I V and Napartovich A P 1993 {\it J. Phys. D: Appl. Phys.} {\bf 26} 418

\bibitem{r7} Dyatko N A, Kochetov I V and Napartovich A P 1992 {\it Sov. J. Plasma Phys.} {\bf 18} 462


\end{thebibliography}
\end{document}